\documentclass[conference]{IEEEtran}
\IEEEoverridecommandlockouts
\usepackage{cite}
\usepackage{graphicx}
\usepackage{amsmath,amssymb,amsfonts}
\usepackage[center]{caption}
\usepackage[font=footnotesize]{caption}
\usepackage{xcolor}
\usepackage{algorithm,algorithmicx,algpseudocode}
\usepackage{bm}
\usepackage{mathtools}
\usepackage{nccmath}
\usepackage{enumitem}
\usepackage{bbm}
\usepackage{multicol}
\usepackage{circuitikz}
\usepackage{verbatim}
\usepackage{import}
\usepackage{array}
\usepackage[font=footnotesize]{caption}

\usepackage{amsthm}

\DeclareMathOperator*{\argmin}{arg\,min}

\title{Optimizing Version AoI in Energy-Harvesting IoT: Model-Based and Learning-Based Approaches}
\author{
		\IEEEauthorblockN{Erfan Delfani and Nikolaos Pappas\\} 
		\IEEEauthorblockA{Department of Computer and Information Science, Link\"{o}ping University, Link\"{o}ping, Sweden} 
        Email: \{erfan.delfani, nikolaos.pappas\}@liu.se
        \thanks{This work has been supported in part by the Swedish Research Council (VR), ELLIIT, and the European Union (ELIXIRION GA 101120135, ETHER GA 101096526, and SOVEREIGN GA 101131481).}
	}

\begin{document}

\maketitle

\begin{abstract}
    Efficient data transmission in resource-constrained Internet of Things (IoT) systems requires semantics-aware management that maximizes the delivery of timely and informative data. This paper investigates the optimization of the semantic metric Version Age of Information (VAoI) in a status update system comprising an energy-harvesting (EH) sensor and a destination monitoring node. We consider three levels of knowledge about the system model—fully known, partially known, and unknown—and propose corresponding optimization strategies: model-based, estimation-based, and model-free methods. By employing Markov Decision Process (MDP) and Reinforcement Learning (RL) frameworks, we analyze performance trade-offs under varying degrees of model information. Our findings provide guidance for designing efficient and adaptive semantics-aware policies in both known and unknown IoT environments.
\end{abstract}

\section{Introduction}

In the rapidly expanding IoT landscape, the ability to transmit all required data is crucial for ensuring effective monitoring and informed decision-making. However, the communication and processing resources needed to manage this data are often limited, particularly as IoT devices—such as those used in healthcare or industrial settings—generate data at an increasing rate. Consequently, it is neither practical nor necessary to transmit all data equally. This constraint has led to growing interest in semantics-aware data management, which considers the freshness, relevance, and usefulness of information to the receiving system. These semantic attributes help maximize the transmission and processing of only the important data, thereby enabling more efficient use of limited resources and enhancing overall system performance.

To address these aspects more effectively, various semantic metrics have been proposed, including Age of Information (AoI)\cite{kaul2012real}, Age of Incorrect Information (AoII)\cite{maatouk2020age}, and Version Age of Information (VAoI)~\cite{yates2021age}. Each metric offers a distinct perspective for evaluating the significance of information over time. AoI quantifies the staleness of information, whereas AoII incorporates a distortion-aware dimension by penalizing updates that are both outdated and incorrect—that is, when the receiver’s content differs from the source. However, in many IoT scenarios—characterized by the impracticality of fully knowing the exact content of the source—VAoI presents a balanced and efficient alternative. By monitoring how \emph{versions} of a monitored phenomenon are delivered in a timely manner, VAoI captures the loss of version changes while minimizing dependence on the exact content of the information source, thereby making it well-suited for real-world IoT scenarios.

In various IoT applications—such as environmental monitoring, smart grids, industrial automation, healthcare, and intelligent transportation systems—semantic metrics like VAoI enable the prioritization of updates based on their contextual relevance rather than solely on their recency or frequency. This prioritization is critical in energy-constrained environments, where minimizing unnecessary transmissions can significantly prolong device lifespan and alleviate network congestion. In specialized domains like the Internet of Medical Things (IoMT), VAoI is crucial for applications including wearable ECG monitoring, glucose tracking, remote fall risk assessment, and ICU sepsis detection. In these cases, timely and relevant updates have a direct impact on patient safety and clinical decision-making. By managing data transmission according to semantic importance, VAoI-driven strategies improve both the responsiveness and efficiency of IoT systems across general and specialized contexts.

A key factor affecting the efficacy of semantics-aware management is the extent of knowledge available about the information source and the communication channel. Depending on the scenario, this knowledge may be complete, partial, or absent. Naturally, more accurate models—made possible by more comprehensive system knowledge—enable more effective performance optimization and yield more reliable outcomes. However, in many practical settings, complete knowledge may not be attainable due to system complexity or inherent dynamics. This paper focuses on optimizing the semantic metric VAoI within a status update system comprising an EH sensor and a monitoring destination node. We consider three progressively informed scenarios regarding knowledge of the source and channel: (i) a fully known model with known parameters, (ii) a known model with unknown parameters, and (iii) a completely model-free environment. Correspondingly, we propose and analyze three optimization strategies: model-based, estimation-based, and model-free approaches. Specifically, we employ MDP and RL frameworks, highlighting the trade-offs and performance differences that arise from operating under different levels of model knowledge. This study offers valuable insights for designing efficient, semantics-aware IoT data management policies that adapt effectively to uncertainty in real-world applications.

\section{Related Works}
Recent studies have explored the use of MDP and RL to optimize AoI and other semantic metrics in EH communication systems.
In \cite{delfani2024state}, the authors investigate an EH sensor that monitors a source with two macro states: normal and alarm. They optimize a state-aware cost function comprising two AoI terms with different exponents, reflecting the increased urgency associated with the alarm state. An MDP problem is formulated, and the optimal policy is derived using Value Iteration.
In \cite{abd2020reinforcement}, an RF-powered IoT device samples and transmits updates over a fading channel. The MDP formulation captures the battery dynamics, AoI, and channel state, with the objective of minimizing the average AoI through an optimal online policy. The study presents the structural properties of the policy and provides performance comparisons with throughput-optimal policies.
In addition, it extends the system to multiple sources, aiming to minimize the long-term average weighted AoI. A Deep Reinforcement Learning (Deep RL) algorithm is proposed to facilitate efficient policy learning.
\cite{ceran2021reinforcement} considers an EH transmitter over an error-prone channel, deciding among sampling, retransmitting, or idling. An average-cost MDP accounts for energy constraints, and age-optimal policies are obtained via Relative Value Iteration and RL for known and unknown system statistics, respectively.
In \cite{gindullina2021age}, an EH node with a finite battery receives updates from heterogeneous sources. An MDP is formulated to minimize average AoI by optimally selecting update requests or idling. 
\cite{jaiswal2021minimization} analyzes time-averaged AoI in EH remote sensing over time-varying channels. The MDP-derived optimal sampling policies are threshold-based, factoring in energy, process age, and channel state.
\cite{crosara2021stochastic} models an EH-powered ARQ system with an MDP to balance retransmissions and new updates under varying energy costs, minimizing AoI for successfully delivered packets.
\cite{hatami2022demand} addresses on-demand AoI minimization under energy and transmission constraints in a multi-user EH IoT network. An MDP-based iterative algorithm is proposed, along with a low-complexity alternative for large-scale systems.
In \cite{hatami2021aoi}, the same problem is studied without transmission constraints. A model-free Q-learning approach is used to learn optimal policies when system dynamics are unknown. \cite{holm2021freshness} introduces a pull-based model where AoI is assessed only at query times, using the QAoI metric in an MDP framework. The optimal policy is derived for periodic queries to an EH sensor. 
The authors in~\cite{zakeri2024semantic} study semantic-aware sampling and transmission for energy harvesting tracking systems with partial observability, using a Partially Observable MDP (POMDP) framework to optimize AoII and distortion metrics via reinforcement learning. 

While prior works have not considered VAoI as a performance metric, some recent studies have focused on optimizing VAoI under the assumption of fully known source and channel models. These include status update systems with EH sensors in gossiping networks \cite{delfani2024version,kaswan2025age}, satellite networks \cite{delfani2025LeoSats}, and query-based setups \cite{delfani2024qvaoi}; as well as scheduling in fading broadcast channels \cite{karevvanavar2024version}, federated learning \cite{hu2024version}, and remote tracking of Markovian sources \cite{salimnejad2024age}. In contrast, our work explicitly adopts VAoI as the key performance metric and contributes by addressing scenarios with unknown source and channel models. We extend existing methods to operate effectively under these more realistic and challenging conditions.

\section{System Model}
\label{Sec_SysModel}

\begin{figure}[t]
	\centering
	\includegraphics[scale=0.084]{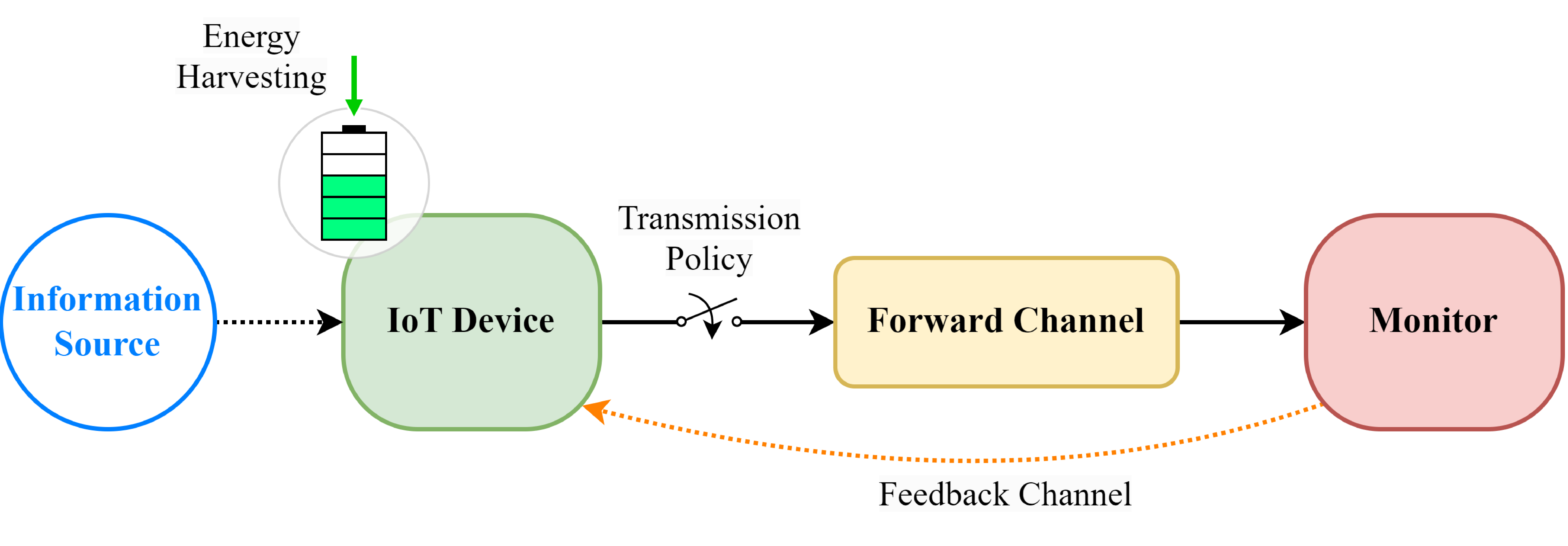} 
    \caption{The status update system model.}
    \label{Fig_SystemModel}
\end{figure}

We consider an end-to-end status update system, as illustrated in Fig. \ref{Fig_SystemModel}, in which data packets sampled from an information source, are transmitted from an IoT device to a destination node via an unreliable communication channel. The transmitting device harvests energy from ambient sources, such as sunlight, and stores it in a rechargeable battery with a finite capacity of $B$ units. The primary source of energy consumption in the device is the transmission action, and we assume that each transmission consumes one battery unit. Our objective is to design a \emph{transmission policy} for the IoT device that minimizes the time-averaged VAoI for the delivered data packets while satisfying the energy constraints imposed by the harvesting process and the battery capacity. We assume a discrete-time system with time slots indexed by $t \in \{0,1,2,\ldots\}$; each transmission occupies one time slot.

\emph{Energy Harvesting Process:} Energy harvesting from ambient sources, depending on the specific source and environmental conditions, is generally a highly dynamic and complex process. In this context, we consider a general and simple, yet efficient, stochastic model suitable for capturing the long-term behavior of the harvesting process. We assume that each battery unit is charged in one time slot with probability $\beta$, following a Bernoulli process. Let $b_t \in \{0,1,\dots,B\}$ denote the battery state at time $t$, $e_t \in \{0,1\}$ represent the energy arrival process, and $a_t \in \{0,1\}$ indicate the transmission action ($a_t = 1$ for transmission and $0$ for idling). The battery evolution over time is governed by:
\begin{align}
    \label{Eq_Battery_Evolution}
    b_{t+1} = \min \left\{ b_t + e_t - a_t, B \right\},
\end{align}
where the energy arrival process is defined as:
\begin{align}
    e_t = 
    \begin{cases}
        1, & \text{with probability } \beta, \\
        0, & \text{with probability } 1 - \beta.
    \end{cases}
\end{align}
The transmission decision $a_t$ is determined by the adopted policy at the IoT device. Note that, when the battery is empty ($b_t = 0$), transmission is not possible, implying $a_t = 0$.

\emph{Communication Channels:} The forward channel from the device to the monitoring node is assumed to be unreliable; not all transmitted packets are successfully delivered. The probability of successful transmission in a given time slot is denoted by $p_s$. Let $h_t \in \{0,1\}$ represent the success indicator of the forward channel at time $t$, where $h_t = 1$ with probability $p_s$ and $0$ otherwise. Additionally, we assume the presence of a backward channel from the monitoring node to the device, which provides acknowledgment (ACK) feedback upon successful reception of a packet.

\emph{Version Age of Information (VAoI):} VAoI is defined as the difference between the current version of the source and the last version received at the destination. Specifically, 
\[
\Delta_t = V^S_t - V^D_t,
\]
where $\Delta_t$ denotes the VAoI at time $t$, $V^S_t$ is the version index at the source, and $V^D_t$ is the most recent version successfully received at the destination. The version generation process at the source, denoted by $g_t \in \{0,1\}$, is modeled as a Bernoulli process: in each time slot, a new version is generated, i.e., $g_t=1$, with probability $p_g$; leading to a geometric distribution for inter-generation times. 
Thus, VAoI evolves as follows:
\begin{align}
    \label{Eq_VAoI_Evolution}
    \Delta_{t+1} = 
    \begin{cases}
        \Delta_t + g_t, & a_th_t = 0, \\
        g_t, & a_th_t = 1.
    \end{cases}
\end{align}

This implies that a successful transmission resets the VAoI to either $0$ or $1$, depending on whether a new version is generated during the time slot. If the transmission fails or does not occur, the VAoI increases by either $0$ or $1$, again depending on the version generation outcome. In practice, excessively stale information (i.e., very large VAoI values exceeding a threshold $\Delta_{\text{max}}$) may be considered unnecessary. Therefore, the VAoI can be truncated at high values to simplify analysis or reduce computational complexity.

In what follows, we investigate the optimization of the long-term average of VAoI to derive the optimal transmission policy at the device for scenarios in which the models or model parameters of the information source and channel are either known or unknown. The optimization problem is formulated as follows:
\begin{align}
    \label{Eq_VAoIOptProblem}
    \bar{\Delta} = \min_{\pi \in \Pi} \lim_{T \rightarrow \infty} \frac{1}{T} E\left[ \sum_{t=0}^{T-1} \Delta_t \right].
\end{align}
where $\Pi$ denotes the set of all feasible policies for selecting actions $a_t \in \{0,1\}$, for $t = 0,1,2,\dots$. Solving this optimization problem yields the optimal transmission policy, which addresses the question: \emph{by how many versions the receiver will lag behind the information source on average under the best-performing policy}.

\section{Optimization Scenarios}
Depending on the knowledge that the optimizer possesses regarding the version generation process $g_t$ and its associated parameter $p_g$, as well as the channel performance process $h_t$ and its associated parameter $p_s$, the approach used to solve the optimization problem varies. In all cases, for a given transmission policy, the stochastic process of VAoI $\Delta_t$ evolves as a first-order Markov chain, since its next state depends only on the current state of the process. This allows us to utilize Markov Decision Process (MDP) and Reinforcement Learning (RL) frameworks for the optimization problem. 

\subsection{Fully Known Models}
\label{Sec_FullyKnownMDP}
First, we assume that the version generation and channel performance processes, along with their parameters, are fully known. Under this assumption, the optimization problem \eqref{Eq_VAoIOptProblem} can be directly formulated as a standard \emph{infinite horizon average cost MDP} problem characterized by \emph{states}, \emph{actions}, \emph{transition probabilities}, and a \emph{cost function} \cite{bertsekas2011dynamic}. The state vector is defined as $s_t = [\Delta_t,b_t] \in S$, where $S = \left\{ (\Delta,b) \mid \Delta \in \{0,1,2,\cdots,\Delta_{\text{max}}\},\ b \in \{0,1,2,\cdots,B\} \right \}$ denotes the state space; the action variable is $a_t \in A$, where $A = \{0,1\}$ represents the action space; the transition probabilities $\mathbb{P}\!\left(s_{t+1} \mid s_t,a_t\right) = \mathbb{P}\!\left(\Delta_{t+1} \mid \Delta_t, a_t\right)\mathbb{P}\!\left(b_{t+1} \mid b_t, a_t\right)$ are directly obtained from \eqref{Eq_Battery_Evolution} and \eqref{Eq_VAoI_Evolution}; and the transition cost function is equal to the VAoI after the action, and is expressed as $C(s_t,a_t,s_{t+1}) = \Delta_{t+1}$, or simply $C_t = \Delta_{t+1}$.

Given that the state space and the action space are finite and that the cost function is upper and lower bounded, it can be shown \cite[Sec. 4.2]{bertsekas2011dynamic} that the MDP problem admits an optimal stationary policy independent of time and the initial state of the system. The optimal policy $\pi^\ast$ is a function of the current state of the system, i.e., $s=(\Delta,b)$, and can be obtained by the Bellman equations as follows:
\begin{gather}
    \label{Eq_BellmanObjective}
    \bar{\Delta}^\ast = \min_{a \in \{0,1\}} Q(s,a) \\
	\pi^\ast(s) =  \argmin_{a \in \{0,1\}} Q(s,a), 
\end{gather}
where
\begin{align}
    Q(s,a) = C(s,a)+\sum_{s^\prime\in S}{\mathbb{P}\left(s^\prime \big | s,a\right) V(s^\prime)},
\end{align}
where $V(s)$ denotes the value function, and $Q(s,a)$ denotes the action-value function of the MDP problem. Here, $C(s,a)$ is the average cost per slot, defined by the transition costs as:
	\begin{equation}
		\label{AvgCost_eqn}
		C(s,a) = \sum_{s^\prime\in \mathcal{S}} {\mathbb{P}\left(s^\prime \big | s,a\right) C\left(s,a,s^\prime\right)}, 
	\end{equation}
where $C(s,a,s^\prime) = \Delta^\prime$.
    
The value function does not have a closed-form expression; however, it can be obtained using iterative dynamic programming approaches such as the Relative Value Iteration Algorithm (RVIA), which yields the optimal policy, $\pi^\ast$~\cite{bertsekas2011dynamic}.

\subsection{Known Model with Unknown Parameters}
\label{Sec_UnkParams}
In this scenario, it is assumed that the stochastic models of the processes $g_t$ and $h_t$ follow Bernoulli distributions, while their parameters, namely $p_g$ and $p_s$, remain unknown. These parameters have physical interpretations and can be estimated using well-established model-based estimation methods, such as the Maximum Likelihood Estimation (MLE). It can be easily shown that the ML estimates of the probabilities $p_g$ and $p_s$ in the Bernoulli processes $g_t$ and $h_t$ are equal to the average number of occurrences of these events up to the current time $t$ \cite[Sec. 3.3]{murphy2012machine}:
\begin{align}
    \hat{p}_g(t) = \frac{1}{t} \sum_{k=0}^{t-1} g_k, \\
    \hat{p}_s(t) = \frac{1}{t} \sum_{k=0}^{t-1} h_k,
\end{align}
where these estimates can also be expressed in a recursive form:
\begin{align}
    \hat{p}_g(t) = \frac{1}{t} g_{t-1} + \frac{t-1}{t} \hat{p}_g(t-1), \\
    \hat{p}_s(t) = \frac{1}{t} h_{t-1} + \frac{t-1}{t} \hat{p}_s(t-1),
\end{align}

The estimation becomes increasingly precise over time, and by using the estimated parameters, the system can be regarded as a fully known setup, as described in the Section \ref{Sec_FullyKnownMDP}, allowing the optimization to be performed accordingly. This results in an \emph{estimation-based MDP} approach.

\subsection{Unknown Models}
\label{Sec_UnkModels}
When the stochastic model of the version generation and channel performance processes are unknown, model-free RL approaches can be employed to solve the optimization problem by learning the system dynamics. RL enables an agent to learn optimal actions by interacting with the system’s environment and receiving cost feedback, making it particularly suitable for problems with unknown or complex dynamics. Among RL techniques, Q-learning, a model-free algorithm, is well-suited for average-cost infinite-horizon MDPs \cite{bertsekas2011dynamic,sutton2018reinforcement}. It iteratively updates estimates of the action-value function based on observed transitions and incurred costs, allowing the solution of optimization problems without prior knowledge of the system’s transition probabilities. Learning is organized into episodes indexed by $k = 0, 1, 2, \ldots$. Each episode consists of a sequence of time slots indexed by $t = 0, 1, 2, \ldots$, during which the system transitions between states. At the beginning of each episode, the system is initialized in an initial state $s_0^{(k)}$ that may vary across episodes.

Q-learning aims to estimate the optimal action-value function $Q^*(s,a)$ by minimizing the temporal difference (TD) error, which quantifies the discrepancy between the current Q-value estimate and a one-step lookahead estimate~\cite[Sec. 6.5]{sutton2018reinforcement}. At each time slot $t$ within an episode, the TD error is computed as:

\begin{align}
\label{Eq_TDerror}
\delta_t = C_t - \hat{\lambda}_t + \min_{a' \in \{0,1\}} Q(s_{t+1}, a') - Q(s_t, a_t),
\end{align}
where $C_t$ is the immediate cost incurred at time $t$, and $\hat{\lambda}_t$ is the current estimate of the optimal average cost.

The Q-learning update uses this TD error as follows:
\begin{align}
Q(s_t, a_t) \leftarrow Q(s_t, a_t) + \alpha_t \, \delta_t,
\end{align}
where $\alpha_t$ is the learning rate at time $t$. To guarantee convergence, $\alpha_t$ must satisfy the standard stochastic approximation conditions: $\sum_{t=0}^\infty \alpha_t$ is infinite, and $\sum_{t=0}^\infty \alpha_t^2$ is finite~\cite[Sec. 2.4]{sutton2018reinforcement}.
Common schedules for $\alpha_t$ include:
\begin{itemize}
  \item $\alpha_t = \frac{1}{t+1}$,
  \item $\alpha_t = \frac{1}{\left(N(s_t,a_t)\right)^\omega}$ with $\omega \in (0.5, 1]$, where $N(s_t,a_t)$ is the visit count of $(s_t,a_t)$.
\end{itemize}

Because the true average cost $\bar{\Delta}^*$ is unknown, its estimate $\hat{\lambda}_t$ is updated iteratively, often at a reference state $s_{\text{ref}}$, via
\begin{align}
\hat{\lambda}_{t+1} = \hat{\lambda}_t + \gamma_t \, \delta_t,
\end{align}
where $\gamma_t$ is a small learning rate satisfying similar conditions as $\alpha_t$.

Balancing \emph{exploration} and \emph{exploitation} is essential: exploration involves selecting actions at random to discover their impact on future costs, while exploitation chooses actions believed to minimize cost based on current knowledge. This trade-off is managed by an $\varepsilon$-greedy policy~\cite[Sec. 2.2]{sutton2018reinforcement}, where at time $t$ in episode $k$:
\begin{align}
a_t = 
\begin{cases}
\text{random action}, & \text{with probability } \varepsilon_k, \\
\argmin_a Q(s_t, a), & \text{with probability } 1 - \varepsilon_k.
\end{cases}
\end{align}
Here, $\varepsilon_k$ is the exploration probability that depends on the episode index $k$.

To encourage convergence from exploration to exploitation, $\varepsilon_k$ typically decays over episodes using one of the following schedules:
\begin{itemize}
  \item Polynomial decay: $\varepsilon_k = \frac{\varepsilon_0}{1 + \mu k}$,
  \item Exponential decay: $\varepsilon_k = \varepsilon_0 e^{-\mu k}$,
  \item Inverse square root decay: $\varepsilon_k = \frac{\varepsilon_0}{\sqrt{k+1}}$,
\end{itemize}
where $\varepsilon_0$ and $\mu$ are positive constants.

After sufficient learning, the optimal policy is given by:
\begin{align}
\pi^*(s) = \argmin_{a \in \{0,1\}} Q^*(s,a).
\end{align}

\section{Numerical Results}
This section presents a simulation-based evaluation of the performance of the MDP, estimation-based MDP, and RL approaches in optimizing the VAoI in the system.

\subsection{VAoI with Fully Known Models}
\begin{figure}[t]
	\centering
	\includegraphics[scale=0.5]{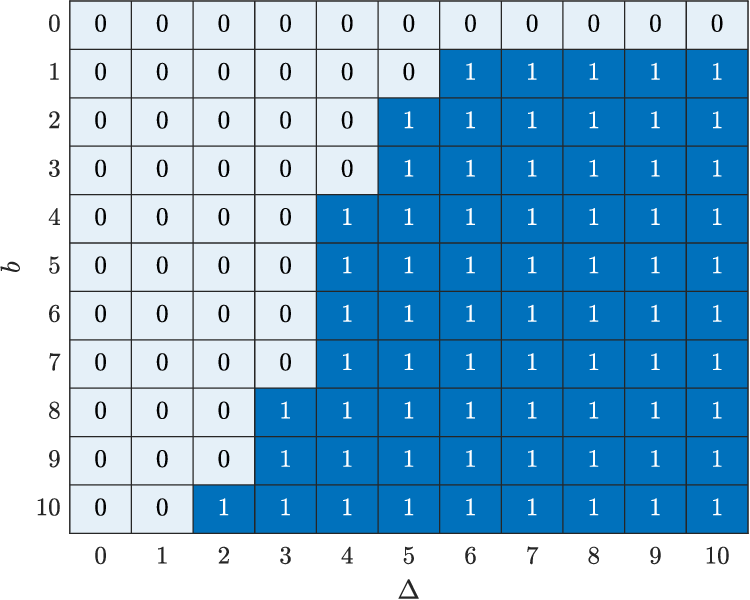} 
    \caption{Optimal Actions for each state $(\Delta,b)$.}
    \label{Fig_OptimalActions}
\end{figure}

We solve the Bellman equation \eqref{Eq_BellmanObjective} using the RVIA algorithm and obtain the optimal policy. The average VAoI is then computed by averaging over $1000$ Monte Carlo iterations, each spanning $10000$ time slots. The optimal action for each state $(\Delta,b) \in S$ is depicted in Fig. \ref{Fig_OptimalActions} for $p_g=0.3$, $p_s=0.8$, $\beta=0.1$, $B=10$, and $\Delta_{\text{max}}=10$. As observed, the optimal policy exhibits a threshold-based structure, wherein for each battery state $b$, the optimal action switches from $0$ to $1$ as the VAoI $\Delta$ exceeds a certain threshold. This is a key characteristic of semantics-aware communication policies, which prevent early battery depletion by avoiding unnecessary energy consumption when the VAoI is low. Instead, transmissions are postponed until the VAoI increases, indicating that updates are genuinely necessary.

\begin{figure}[t]
	\centering
	\includegraphics[scale=0.45]{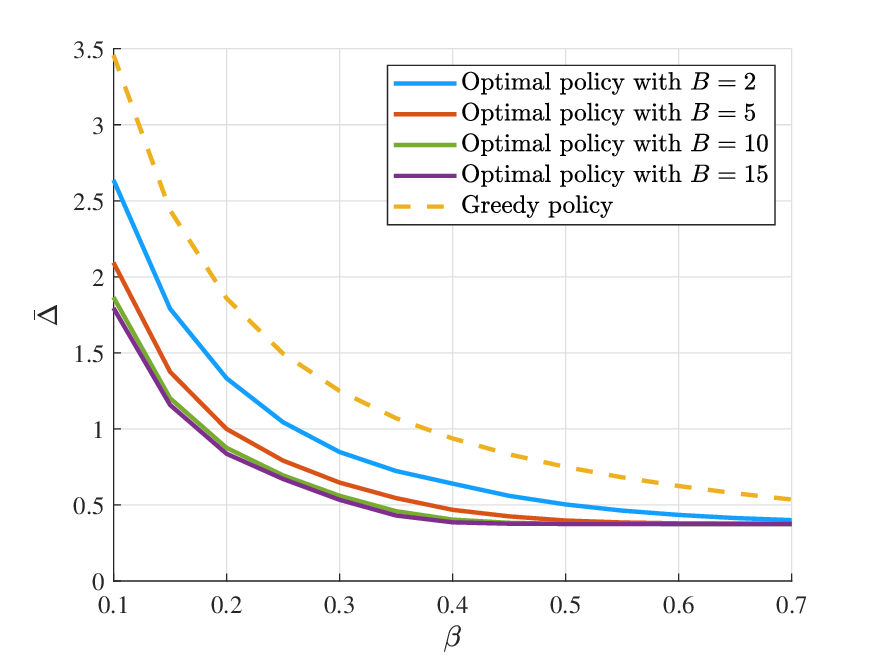} 
    \caption{Optimal average VAoI vs. $\beta$.}
    \label{Fig_OptimalVAoIvsBeta}
\end{figure}

For the fully known model, the average VAoI under the optimal policy is shown in Fig. \ref{Fig_OptimalVAoIvsBeta} as a function of $\beta$ for various battery sizes $B$. As expected, an increase in the energy arrival probability $\beta$ improves the VAoI at the monitoring node. Similarly, an increase in the battery size $B$ also enhances (i.e., reduces) the VAoI. However, for exceedingly large battery sizes, this improvement becomes negligible, as observed in the Fig. \ref{Fig_OptimalVAoIvsBeta} for $B=10$ to $B=15$.

As a baseline policy, we also consider a myopic greedy policy in which a transmission occurs as soon as energy becomes available, i.e., the battery state transitions from $0$ to $1$ upon energy arrival. For the greedy policy, a battery size larger than $1$ offers no benefit, as the battery charge is consumed greedily and the battery level never exceeds $1$. The average VAoI for the greedy policy is depicted in Fig. \ref{Fig_OptimalVAoIvsBeta}, where the superior performance of the optimal policy becomes more pronounced as the energy arrival probability $\beta$ decreases, i.e., when the system becomes more energy-constrained. In such cases, adopting a semantics-aware policy becomes increasingly critical for achieving satisfactory performance.

\emph{The effect of $\Delta_{\text{max}}$:} As mentioned in Section \ref{Sec_SysModel}, storing excessively high VAoI values (i.e., those exceeding $\Delta_{\text{max}}$) may be unnecessary; therefore, the VAoI variable can be truncated at this maximum value. Table \ref{Table_VAoI_Dmax} presents the average VAoI obtained from the optimal policy for different values of $\Delta_{\text{max}}$, with $\beta=0.1$. As observed, truncating the VAoI at values greater than $10$ or $15$ does not cause significant differences, as the impact is negligible. Accordingly, we select $\Delta_{\text{max}}=10$ for the remaining of the numerical results in this section.

\renewcommand{\arraystretch}{1.5}
\begin{table}
    \centering
    \caption{Optimal average VAoI as a function of $\Delta_{\text{max}}$.}
    \begin{tabular}{|m{0.45cm}|m{0.4cm}|m{0.4cm}|m{0.4cm}|m{0.4cm}|m{0.4cm}|m{0.4cm}|m{0.4cm}|m{0.4cm}|m{0.4cm}|}
        \hline
         $\Delta_{\text{max}}$ & 1 & 2 & 3 & 4 & 5 & 10 & 15 & 20 & 25\\
         \hline
         $\bar{\Delta}$ & 0.81 & 1.36 & 1.64 & 1.74 & 1.76 & 1.79 & 1.80 & 1.80 & 1.80\\
         \hline
    \end{tabular}
    \label{Table_VAoI_Dmax}
\end{table}

\subsection{VAoI with Unknown Parameters and Unknown Models}
\begin{figure}[t]
	\centering
	\includegraphics[scale=0.45]{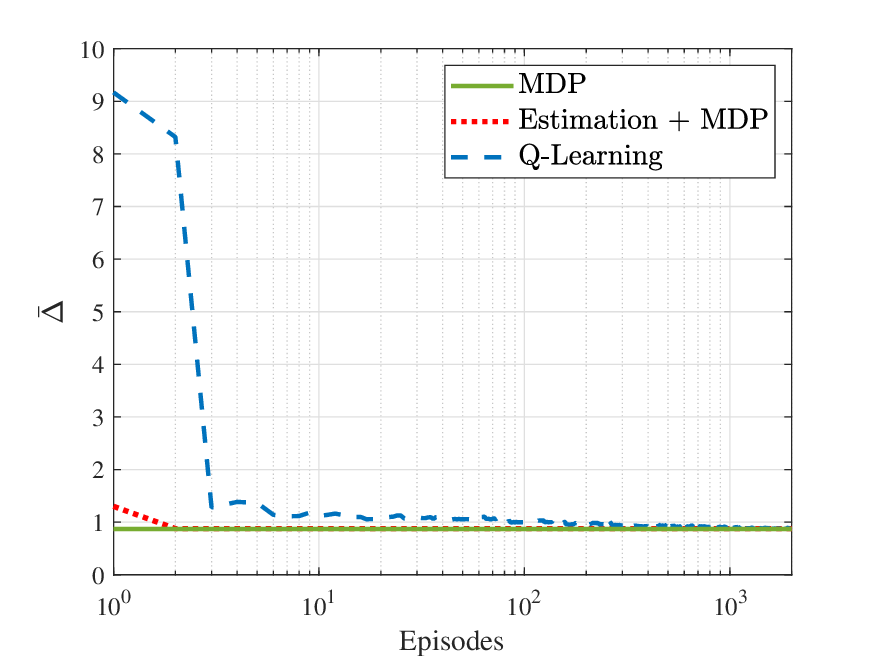} 
    \caption{Average VAoI for different approaches with $\beta=0.2$.}
    \label{Fig_VAoIepisodesBeta02}
\end{figure}

\begin{figure}[t]
	\centering
	\includegraphics[scale=0.45]{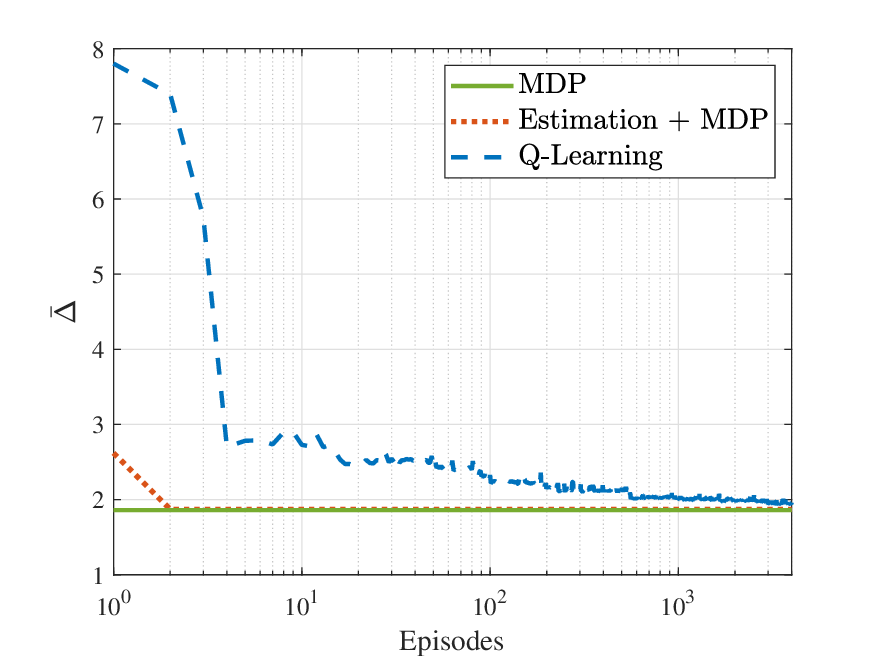} 
    \caption{Average VAoI for different approaches with $\beta=0.1$.}
    \label{Fig_VAoIepisodesBeta01}
\end{figure}

For unknown parameters and models, estimation-based optimization of the MDP, as discussed in Section \ref{Sec_UnkParams}, and Q-learning approaches, as described in Section \ref{Sec_UnkModels}, are utilized, respectively. Both approaches are simulated over episodes of $2000$ time slots. After each episode, the obtained policy is used to calculate the average VAoI for both approaches, with the evolution of the average VAoI over episodes depicted in Figs. \ref{Fig_VAoIepisodesBeta02} and \ref{Fig_VAoIepisodesBeta01} for $\beta=0.2$ and $\beta=0.1$, respectively. The optimal average VAoI for the fully known MDP is also presented as a reference. For the Q-learning algorithm, we used the exploration probability $\varepsilon_k=\max \{\frac{1}{\sqrt{k+1}},0.1\}$, learning rate $\alpha_t=\frac{1}{(N(s_t,a_t))^{0.55}}$, and $\gamma_t=\frac{1}{(N(s_t,a_t))^{0.6}}$. Random initial states $s^{k}_0$ were selected across the state space, with $\Delta_{\text{max}}=10$ and $B=10$.

As observed in Figs. \ref{Fig_VAoIepisodesBeta02} and \ref{Fig_VAoIepisodesBeta01}, for both $\beta$ values, the estimates of $p_g$ and $p_s$ rapidly converge to their true values. After only one episode, the optimal policy derived from the estimation-based MDP performs equivalently to the optimal policy obtained from a fully known model. In contrast, Q-learning—which does not rely on any knowledge of the underlying stochastic model governing version generation or channel performance—requires more than a thousand episodes to achieve the optimal VAoI. This is because Q-learning must learn the value of every possible state-action pair, $Q(s,a)$, across the entire state and action spaces ($s \in S$, $a \in A$). To accomplish this, it must repeatedly attempt different actions in various states and observe the resulting costs. The learning process is further hindered by the system’s complex dynamics, which give rise to a large number of possible transitions.

\emph{Remark:} These results demonstrate that possessing knowledge of the system variables—in this case, the version generation and channel performance processes—enables accurate modeling of the system. In such scenarios, the optimal policy can be obtained either in a single run (for fully known models) or after a few runs (for models with known structure but unknown parameters) of dynamic programming. However, when both the model structure and parameters are unknown, learning-based methods such as Q-learning can be employed. In these cases, the action-value function of the system is learned through a data-driven approach involving multiple iterations, potentially a large number. Nevertheless, the RL method also offers lower computational complexity compared to RVIA\footnote{The complexity of RVIA per iteration is $\mathcal{O}(|S|^2|A|)$, while that of Q-learning is $\mathcal{O}(1)$, where $|S| = (\Delta_{\text{max}} + 1)(B + 1)$ and $|A| = 2$ denote the sizes of the state and action spaces, respectively.}, making it a more suitable approach for solving problems with large state spaces.

\section{Conclusion}
In this work, we optimized the VAoI in a system for status updating from an EH sensor to a monitoring node by employing MDP, estimation-based MDP, and Q-learning. These methods were applied to scenarios involving source and channel models, including a fully known model, a stochastic model with unknown parameters, and a completely unknown model. The results demonstrated that greater knowledge of the system led to more accurate models and faster convergence. However, in more realistic environments with limited information, estimation-based and learning-based approaches still effectively achieved optimal or near-optimal performance. Future research could explore leveraging the threshold-based structure with reinforcement learning to achieve faster and more efficient convergence, as well as investigating the effectiveness of Deep RL in large state spaces.

\bibliographystyle{IEEEtran}
\bibliography{Refs}

\end{document}